\documentclass[fleqn,10pt]{wlscirep}
\usepackage[utf8]{inputenc}
\usepackage[T1]{fontenc}

\usepackage{amssymb}
\usepackage{physics}
\usepackage{float}
\usepackage{graphicx}
\usepackage{sidecap}
\usepackage{tabularx}
\usepackage{array}
\usepackage{booktabs}
\usepackage{tikz}
\usepackage{longtable}
\usepackage{hyperref}
\usepackage{booktabs}
\usepackage{cellspace} 
\usepackage{tabularx} 
\usepackage{float}

\def\MA{\textcolor{black}}

\def\MAr{\textcolor{black}}

\title{\centering Characterization of Nanostructural Imperfections in Superconducting Quantum Circuits}

\author[1, *]{\centering Mohammed Alghadeer} 

\author[1]{Simone D. Fasciati}

\author[1]{Shuxiang Cao}

\author[1]{Michele Piscitelli}

\author[2]{Matthew C. Spink}

\author[2]{David G. Hopkinson}

\author[2]{Mohsen Danaie}

\author[3]{Susannah C. Speller}

\author[1, *]{\\ Peter J. Leek} 

\author[1, *]{Mustafa Bakr}

\affil[1]{Department of Physics, Clarendon Laboratory, University of Oxford, Oxford, OX1 3PU, United Kingdom}
\affil[2]{Diamond Light Source Ltd. electron Physical Science Imaging Centre (ePSIC), Didcot, OX11 0DE, United Kingdom}
\affil[3]{Department of Materials, University of Oxford, Oxford, OX1 3PH, United Kingdom}

\affil[*]{mohammed.alghadeer@physics.ox.ac.uk}
\affil[*]{peter.leek@physics.ox.ac.uk}
\affil[*]{mustafa.bakr@physics.ox.ac.uk}

\begin{abstract}
Decoherence in superconducting quantum circuits, caused by loss mechanisms like material imperfections and two-level system (TLS) defects, remains a major obstacle to improving the performance of quantum devices. In this work, we present atomic-level characterization of cross-sections of a Josephson junction and a spiral resonator to assess the quality of critical interfaces. Employing scanning transmission electron microscopy (STEM) combined with energy-dispersive X-ray spectroscopy (EDS) and electron-energy loss spectroscopy (EELS), we identify  structural imperfections associated with oxide layer formation and carbon-based contamination, and correlate these imperfections to the \MA{patterning} and etching steps in the fabrication process and environmental exposure. \MA{These results suggest that TLS imperfections at critical interfaces significantly contribute to limiting device performance}, emphasizing the need for an improved fabrication process.

\end{abstract}

\begin{document}

\flushbottom
\maketitle

\section*{Introduction}

Superconducting quantum circuits have emerged as a prominent platform in the pursuit of scalable quantum computing due to their coherence and compatibility with nanofabrication techniques \cite{brecht2015demonstration-1, place2021new-2, vepsalainen2022improving-3}. Recent advancements have seen coherence times extended to several milliseconds \cite{somoroff2023millisecond, ganjam2024surpassing, milul2023superconducting}. However, nanoscopic defects at critical interfaces continue to limit overall performance and uniformity in larger-scale devices \cite{mohseni2024build, zanuz2024mitigating, abdurakhimov2022identification, bilmes2020resolving}. Sources of decoherence are attributed to radiative \cite{houck2008controlling}, dielectric \cite{martinis2005decoherence} and quasiparticle \cite{catelani2011relaxation} decay mechanisms. In particular, loss originating from material imperfections such as two-level system (TLS) defects present at metal-air (MA), metal-substrate (MS), and substrate-air (SA) interfaces \cite{martinis2005decoherence-10, muller2019towards-11, gao2008experimental-12}, play a crucial role in energy loss in both qubits and resonators. These imperfections manifest as fluctuating electromagnetic environments, disrupting qubit coherence and quantum gate operation. Despite extensive research, the detailed mechanisms by which these interface defects contribute to decoherence are not yet fully understood, which highlights the need for a thorough characterization at the atomic scale \cite{muller2019towards-11}.

Efforts to mitigate structural imperfections and TLS loss include modifications to circuit geometries \cite{gao2008experimental-12, woods2019determining-16} and exploring new materials \cite{wisbey2010effect-17, earnest2018substrate-18, mcrae2020dielectric-19, murray2021material-20}. Although the precise origins of TLS loss remain a topic of discussion \cite{muller2019towards-11}, it is predominantly attributed to structural and chemical defects at air interfaces \cite{gao2008physics-21, calusine2018analysis-22}, arising primarily from fabrication-related chemical residues and oxidation due to exposure to ambient environment \cite{muller2019towards-11, woods2019determining-16, lisenfeld2015observation-23}. Developing a deeper understanding of the microscopic mechanisms underlying TLS loss is crucial for improving quantum device fabrication. This requires advanced characterization methods to identify structural and chemical defects directly and to inform more effective mitigation strategies \cite{martinis2005decoherence-10}.

Improvements in the MS interface have been achieved through substrate cleaning \MA{by oxide etching} before thin-film growth \cite{earnest2018substrate-18}, controlled epitaxial film deposition \cite{richardson2016fabrication-24}, and interface-targeted etching techniques for resonator and qubit structures \cite{megrant2012planar-25, richardson2020low-26}. However, controlling native oxide formation at MA and SA interfaces remains challenging, as thin films and substrates are typically exposed to ambient conditions. Chemical etching of oxides at SA and MA interfaces can solve part of the problem and significantly improve resonator quality factors \cite{altoe2022localization-27}, however this is limited by the rapid regrowth of oxide layers. 

Post-fabrication protective techniques, such as self-limiting oxide growth \cite{ding2023stable-28}, nitrogen plasma treatments \cite{zheng2022nitrogen-29}, surface encapsulation with metallic layers \cite{bal2024systematic, chang2024eliminating, karuppannan2024improved} and surface passivation using molecular self-assembled monolayers \cite{alghadeer2022surface-30, alghadeer2024mitigating} offer potential solutions to mitigate the regrowth of oxide layers. Moreover, isolating superconducting circuits from bulk substrates using suspended membranes shows promise in reducing substrate-related loss and correlated errors mediated by quasiparticle propagation \cite{chistolini2024performance, junger2024implementation}. Despite these advancements, most techniques require further refinement to address the practical challenges associated with integrating these into a standard fabrication process. Factors such as process complexity, integration and compatibility with existing fabrication techniques must be carefully considered to make sure that no major side effects further contribute to structural and chemical defects at relevant interfaces. A detailed understanding of how these modifications in the fabrication process affect the properties of interfaces is crucial for developing effective and scalable solutions for fabricating low-loss quantum devices \cite{mcrae2020dielectric-19}.

In this work, we present detailed materials characterization of the metal-substrate (MS), metal-air (MA), and substrate-air (SA) interfaces of a superconducting quantum device fabricated at University of Oxford, focusing on oxide layer formation and carbon contamination. Using high-resolution cross-sectional imaging with scanning transmission electron microscopy (STEM) combined with energy-dispersive X-ray spectroscopy (EDS) and electron energy-loss spectroscopy (EELS), we examine a Josephson junction (JJ) and a spiral resonator. The nanostructural imperfections identified are correlated with the \MA{patterning} and etching steps in the fabrication process and environmental exposure \cite{martinis2005decoherence-10, altoe2022localization-27}. Our study \MA{emphasizes} on the need of using materials characterization techniques for assessing nano-scale surface roughness and interface oxidation, providing critical insights about nanostructural imperfections in superconducting quantum devices.

\section*{Methodology}
Superconducting quantum circuits are commonly fabricated using thin-films of aluminum, niobium, or titanium alloys on silicon or sapphire substrates \cite{In13}. The most critical components of these circuits are Josephson junctions, which when shunted with large capacitors, can form the widely used type of superconducting qubits known as the transmon \cite{CQ1}. The fabrication process of these junctions can vary based on the junction size controlled by electron-beam lithography exposure dose, oxidation parameters, and metal evaporation pressure. In our study, we employ a double-sided fabrication on a double-side 3-inch polished intrinsic silicon wafers, involving multiple lithography steps, thin-film depositions, and protective resists to ensure high-quality surfaces and interfaces on both sides during the fabrication process.

\subsection*{Fabrication Process}
\label{subsec:Fabrication}

\begin{figure}[H]
  \centering
   \includegraphics[width=0.6\textwidth]{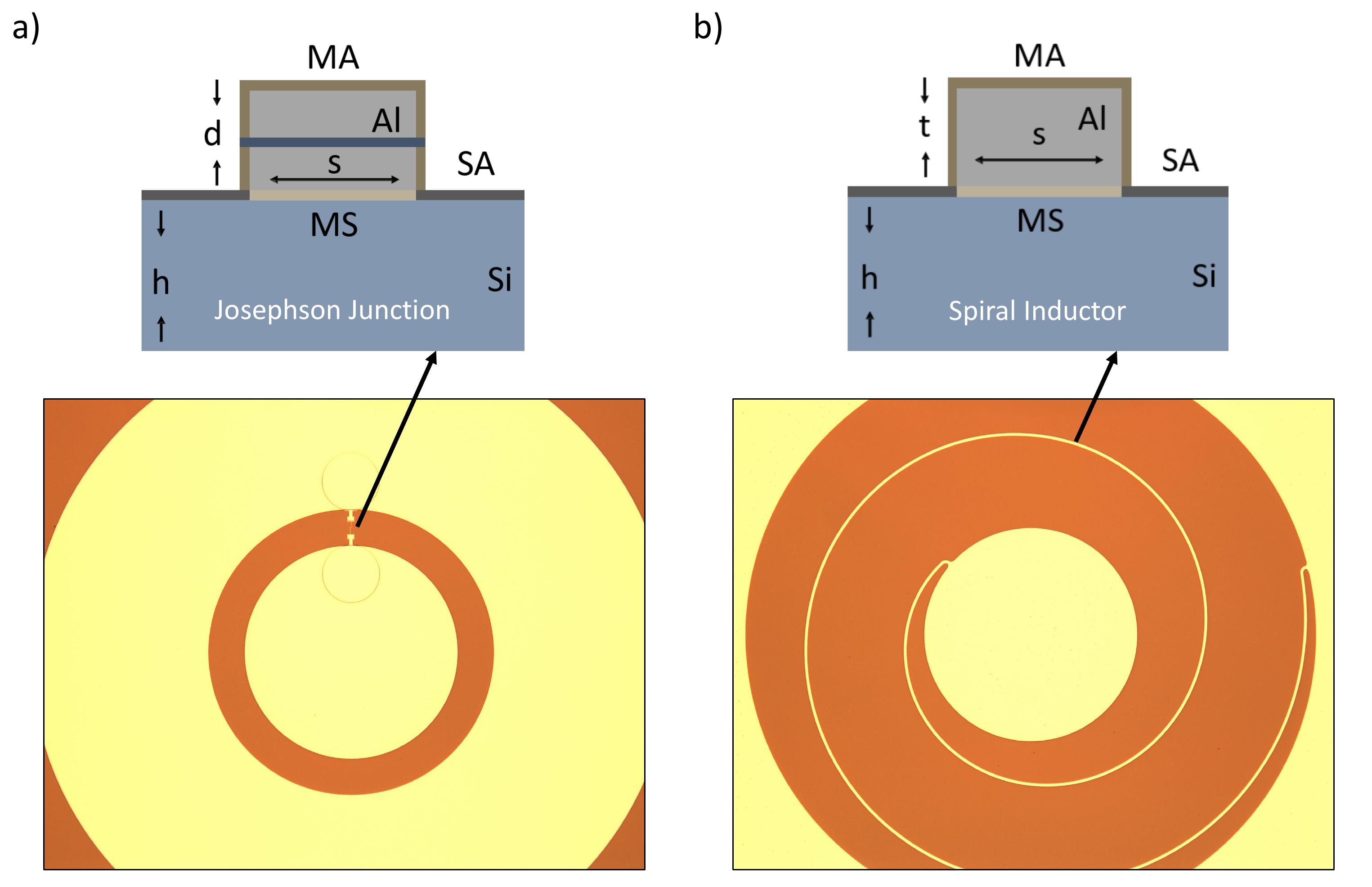}
  \caption{\MAr{(a) Optical image of the qubit indicating the location where the lamella was extracted for scanning transmission electron microscopy (STEM) analysis. (b) Optical image of the resonator showing the lamella extraction site}.}
  \label{F1a}
\end{figure}

\MAr{The superconducting quantum device studied in this work consists of transmon qubits arranged in a \( 4 \times 4 \) lattice within a 3D-integrated coaxial circuit quantum electrodynamics (cQED) architecture \cite{rahamim2017double, spring2022high, fasciati2024complementing}. The qubits and resonators are fabricated on opposite sides of the silicon substrate and capacitively coupled through the bulk substrate. The coupling strength is primarily determined by the substrate thickness and the geometry of the capacitive pads of both the qubit (in Fig.~\ref{F1a}(a)) and the resonator (in Fig.~\ref{F1a}(b)) \cite{rahamim2017double}}. The detailed steps of the fabrication process are described \cite{peterer2016experiments, cao2023implementation}, with relevant design parameters given in Table ~\ref{Table.1}. Spin-coating a protective photoresist layer on the backside is critical in double-sided fabrication process to protect the wafer and prevent additional contamination. The resonators side is patterned first while the qubit side is covered with photoresist, followed by cleaning the photoresist and spin-coating another protective photoresist layer on the resonators side and \MA{patterning} the qubits side.

\begin{table}[H]
  \centering
  \caption{Geometric design parameters of relevant parts of the device shown in Fig.~\ref{F1}.}
  \label{Table.1}
  \begin{tabular}{Sl Sl Sl}
    \hline
    Geometric parameter          &   Symbol    & Standard value                      \\
    \hline
    Spiral line width            & ~~~~~ $s$       &   5 ~~~~~~~~~~~~~~~~~~~~   µm       \\
    Al thin film thickness       & ~~~~~ $t$       &   100   ~~~~~~~~~~~~~~~~   nm       \\
    JJ thin film thickness   & ~~~~~ $d$       &   (27-30)  +  70      ~~   nm       \\
    Si substrate thickness       & ~~~~~ $h$       &   500   ~~~~~~~~~~~~~~~~   µm       \\
    \hline
  \end{tabular}
\end{table}

\subsubsection*{Wafer Cleaning}

The fabrication process starts with cleaning a high-resistivity (\(>\,10\,\mathrm{K\Omega\cdot cm}\)) intrinsic silicon wafer using a 10:1 buffered oxide etch (BOE) solution of hydrofluoric acid and ammonium fluoride to remove native oxides and contaminants. After etching, the wafers are thoroughly rinsed with ultrapure deionized water, dried with nitrogen gas, and promptly transferred (within 5~min) to minimize re-oxidation before thin-film deposition.

\subsubsection*{Aluminum Thin-Film Deposition}

Immediately after water cleaning, the wafer is immediately loaded in Plassys MEB550S2 at ultra-high vacuum (UHV) and is baked up to \MAr{$200\,^\circ\mathrm{C}$} for 10~min. After which a layer of 100 nm of aluminum is deposited at rate of 1 nm/s on the substrate through UHV electron-beam evaporation under controlled temperature and low-pressure conditions, with a base pressure down to $10^{-9}$~mbar and an evaporation pressure of around $10^{-8}$~mbar, ensuring high purity and uniformity of the thin film. The deposition rate and substrate temperature are carefully controlled to ensure smooth thin-film growth for optimal grain structure.

\subsubsection*{Photolithography and Micro-scale Circuit Elements}

A positive photoresist AZ~1514~H is spin-coated onto the wafer and then exposed to ultraviolet light through a chrome photomask that defines the desired circuit patterns. After development with AZ 726 MIF developer solution, the exposed areas of aluminum are revealed for etching. The aluminum is then selectively etched away using a wet etching process to define the circuit elements. An aluminum etchant Alfa Aesar 44581 solution and water are used to achieve anisotropic etching with optimal selectivity to minimize remaining aluminum defects. This step creates the micro-scale features of the circuit, including capacitors, inductors, and coupling interconnects. Immediately after the etching process, residual resist is removed using DMSO.

\subsubsection*{Electron-Beam Lithography and \MAr{Nano-scale} Josephson Junctions}

For the nano-scale features, high-resolution electron-beam lithography (EBL) is used to define the Josephson junctions. The junctions are fabricated using the Dolan bridge technique \cite{dolan1988very}, which involves double-angle evaporation of aluminum to form the tunnel barriers, followed by careful removal of excess aluminum through a lift-off process. A bilayer resist structure is employed, consisting of a copolymer (MA/MMA) and a polymethyl methacrylate (PMMA) layer, to create an undercut profile necessary for the shadow evaporation process. After spin-coating the resist, EBL is carried out in a JEOL system at 100~keV, using aperture Ap4 size 2~nA - 60~$\mathrm{\mu m^2}$ for small features and Ap8 size 100~nA - 300~$\mathrm{\mu m^2}$ for large features, with doses typically around 1500~$\mathrm{\mu C/cm^2}$. Following the exposure, the critical features are then developed using a mixture of IPA/MIBK mixture in a 3:1 ratio.

After EBL patterning, the wafer is loaded into the Plassys MEB550S2. Prior to deposition, an argon (Ar) ion milling is performed for 1~min (voltage 400~V, acceleration voltage 90~V/s, current 15~mA) to remove any residual contaminants and native oxides from the metal and substrate surfaces, ensuring a clean interface for the subsequent aluminum deposition. The first layer of junction is then deposited at an angle of $60^\circ$ from normal incidence, depositing 60~nm of Al at a rate of 0.5~nm/s. Due to the deposition angle, the effective thickness of the deposited film is approximately 27-30~nm. Following the first deposition, an \textit{in situ} controlled static oxidation inside Plassys is performed, typically for 5-10~min at an oxygen pressure of 5-10~mbar, depending on the target junction resistance. This controlled oxidation forms the thin insulating barrier of aluminum oxide essential for the tunnel junction. After pumping back down to UHV conditions, the second layer of aluminum is deposited at normal incidence ($0^\circ$), depositing 70~nm of Al at a rate of 0.5~nm/s, effectively completing the Josephson junction structure. Precise control over the oxidation parameters, such as oxygen pressure and exposure time, is critical to achieve the desired tunnel barrier properties and, consequently, the critical current of the junction \cite{dolan1988very}. Following evaporation, a lift-off process is carried out in a DMSO solution at \MAr{$80\,^\circ\mathrm{C}$} for around 2~hrs and immediately followed by thoroughly rinsing with ultrapure deionized water and drying using nitrogen gas. 

\subsubsection*{Post-Fabrication Cleaning and Packaging}

Finally, the wafer is spin-coated with protective photoresist on both sides for milling and dicing. The chips are then usually packaged into sample holders before being installed into a cryostat for microwave measurements. Handling both sides of the device carefully is important to prevent introducing contaminants or mechanical damage that could add more defects.

\subsection*{Materials Characterization}

\subsubsection*{Sample Preparation using FIB-SEM}
To investigate the structural characteristics of the fabricated devices, cross-sectional samples were prepared using a dual-beam Focused Ion Beam Scanning Electron Microscope (FIB-SEM, JEOL 4700F). The samples were initially mounted onto holders and transferred into the FIB-SEM system. The regions of interest were identified using SEM imaging, and a protective layer of platinum was deposited first by an electron beam and subsequently by ion beam to protect the surface during milling. This protective layer prevents damage to the sample surface from the ion beam during milling. Details of the FIB process is shown for both the Josephson junction (qubit side in Fig.~\ref{F1}(a)) and the spiral inductor (resonator side in Fig.~\ref{F1}(b)).

\subsubsection*{Lamellae Thinning and Vacuum Transfer}
\begin{figure}[H]
  \centering
   \includegraphics[width=1.0\textwidth]{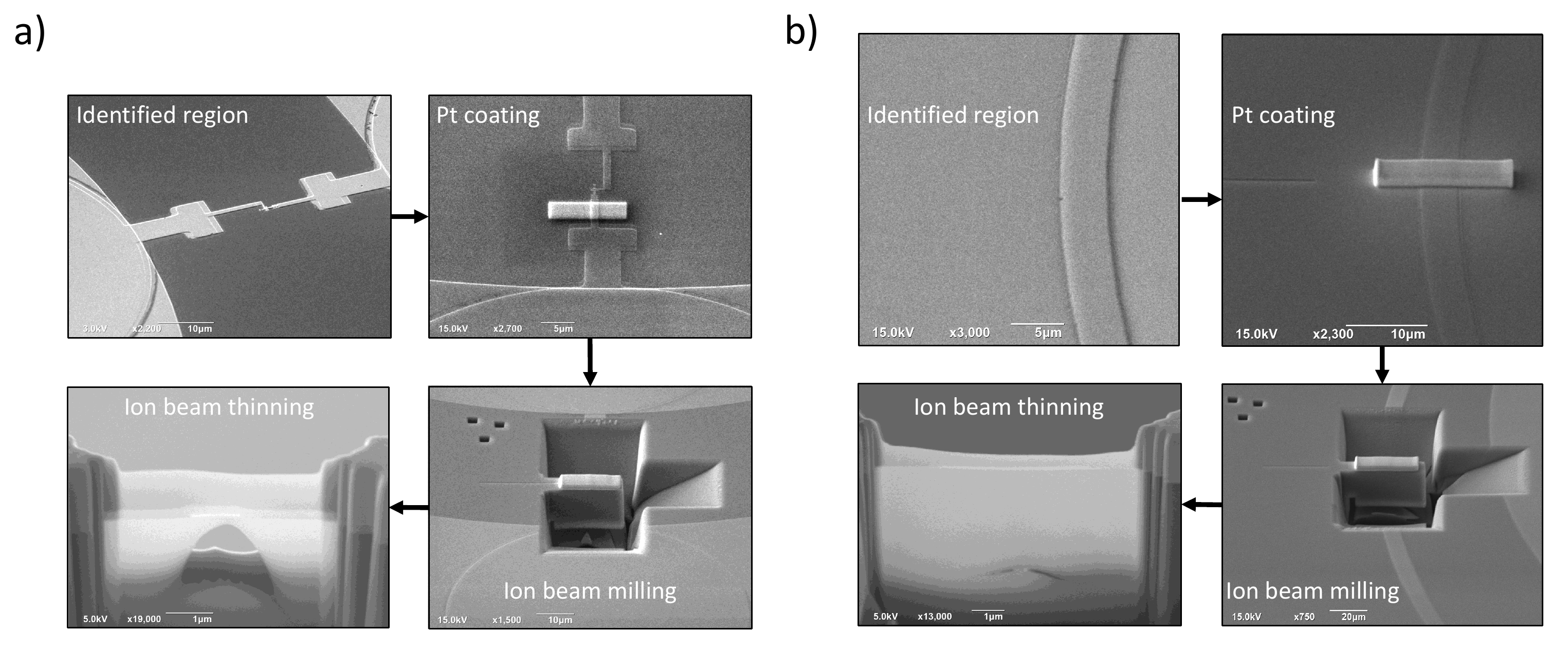}
  \caption{\MAr{Detailed Scanning Electron Microscope (SEM) images during the Focused Ion Beam (FIB) process, illustrating ion beam milling and thinning of the lamella on both (a) the Josephson junction and (b) the spiral inductor. The final fine-thinned lamella for each circuit element is then prepared for high-resolution STEM characterization via controlled vacuum transfer}.}
  \label{F1}
\end{figure}

Cross-sectional lamellae were milled to expose the regions of interest, lifted out, and transferred onto TEM grids \MAr{under a high vacuum of approximately \(3.6 \times 10^{-4}\,~\mathrm{Pa}\).}. These lamellae were then carefully thinned using progressively lower ion beam currents, with final polishing at 10~kV and 5~kV to minimize amorphization and enhance imaging quality shown in Fig.~\ref{F1}(c) and Fig.~\ref{F1}(d). The samples were transferred out of the FIB-SEM under vacuum conditions using a transfer holder and placed in a glovebox to prevent oxidation. This approach ensured that the structural and compositional integrity of the samples was maintained prior to materials analysis. By minimizing exposure to ambient conditions, we minimized the possibility of introducing additional surface oxidation or contamination, which could interfere with the desired characterization.

\subsubsection*{Cross-Sectional Imaging and Spectroscopy Analysis}

Scanning transmission electron microscopy (STEM) was performed at the electron Physical Science Imaging Centre (ePSIC). The qubit side of the device was analyzed with a JEOL ARM300CF aberration-corrected instrument (E02) equipped with an Oxford Instruments XMAX 100 energy-dispersive X-ray spectroscopy (EDS) detector. The resonator side of the device was analyzed with a JEOL ARM200CF aberration-corrected instrument (E01) equipped with a Gatan Dual electron-energy loss spectroscopy (EELS) detector and a large solid angle EDX detector (JEOL Centurio).  The ARM200CF operated with a convergence semi-angle of 23~mrad, an EELS collection semi-angle of 52~mrad and a probe current of 41~pA.  The EELS and EDX data were acquired using the Gatan Microscopy Suite and analyzed using \MAr{Hyperspy software \cite{hyperspy}}. The plasmon background of the EELS were removed by fitting a power law to the region before the oxygen edge in the 450-500~eV range.  The O K-edge, Al K-edge and Si K-edge components in the spectra were fitted using open source generalised oscillator strengths \cite{segger2023software} to compute the double differential cross sections for inelastic electron scattering by atoms. The EDX intensities were obtained using integration windows of width 3~eV, carefully selecting suitable background windows to avoid peak overlaps. EELS and EDX spectra from regions of interest of different size have been normalised by summing the intensity over the region and dividing by its area. Linescans have been created from spectrum images by summing data in a series of 1~nm thick slices with widths indicated on the images. 

\newpage
\section*{Results and Discussion} 
In this section, we present a detailed analysis on two critical components on both sides of the fabricated device: the Josephson junction (qubit side in Fig.~\ref{F1}(a)) and the spiral inductor (resonator side in Fig.~\ref{F1}(b)). \MA{The analysis on oxide thickness, elemental distribution, and surface roughness discussed here represent qualitative trends observed across a limited number of cross-sectional lamellae. Given the inherent non-uniformity of surface oxidation, carbon contamination, and roughness across the device, combined with the localized nature of the TEM-based analysis, a detailed quantitative analysis would not be robust and reproducible. For this reasons, we focus this discussion on spatial trends and the relative differences between interfaces, while including limited quantitative analysis through linescans across the interfaces, rather than stressing on absolute quantitative measures.}

\subsection*{Materials Analysis of a Josephson Junction - Qubit Side}

\begin{figure}[H]
  \centering
   \includegraphics[width=0.7\textwidth]{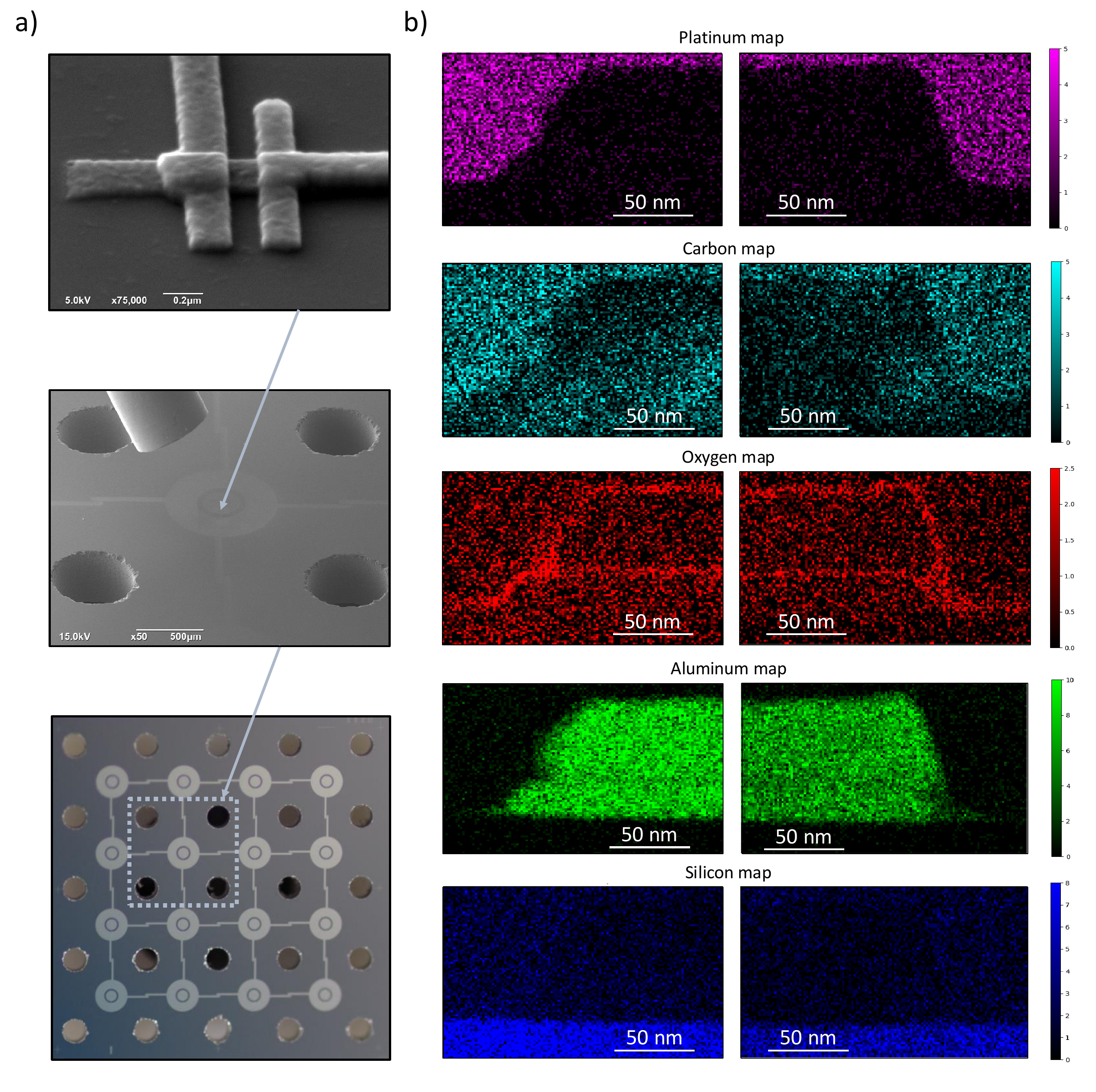}
  \caption{(a) \MA{Scanning electron microscopy (SEM)} images showing the location from which the lamella was extracted from the qubit side for cross-sectional analysis of the Josephson junction. (b) Cross-sectional \MA{STEM combined with dispersive X-ray spectroscopy (EDS)} elemental maps of both sides of the lamella from the Josephson junction area, \MA{highlighting the elemental compositions of oxygen (red), aluminum (green), silicon (blue), carbon (teal), and platinum (purple)}. \MAr{The observed carbon signal within the platinum capping layer originates from the gas injection system (GIS) used during deposition since the precursor is not pure platinum, but Methylcyclopentadienyl trimethyl platinum}.}
  \label{FJJ-1}
\end{figure}

Starting with the Josephson junction sample, Fig.~\ref{FJJ-1}(a) presents SEM images of the device where the lamella was extracted for cross-sectional analysis, and Fig.~\ref{FJJ-1}(b) showing the elemental distribution in the lamella. From Fig.~\ref{FJJ-1}(b), the MS interface is notably clean across the whole sample, exhibiting minimal oxygen content, which suggests a lack of significant oxidation at this interface. This is clear when comparing the MS interface with the pronounced presence of oxygen at MA and SA interfaces as well as at the tunnel barrier between the two evaporated aluminum thin films, corresponding to the intentionally formed AlO\textsubscript{x} layer during the junction oxidation process. 

\begin{figure}[H]
  \centering
   \includegraphics[width=0.95\textwidth]{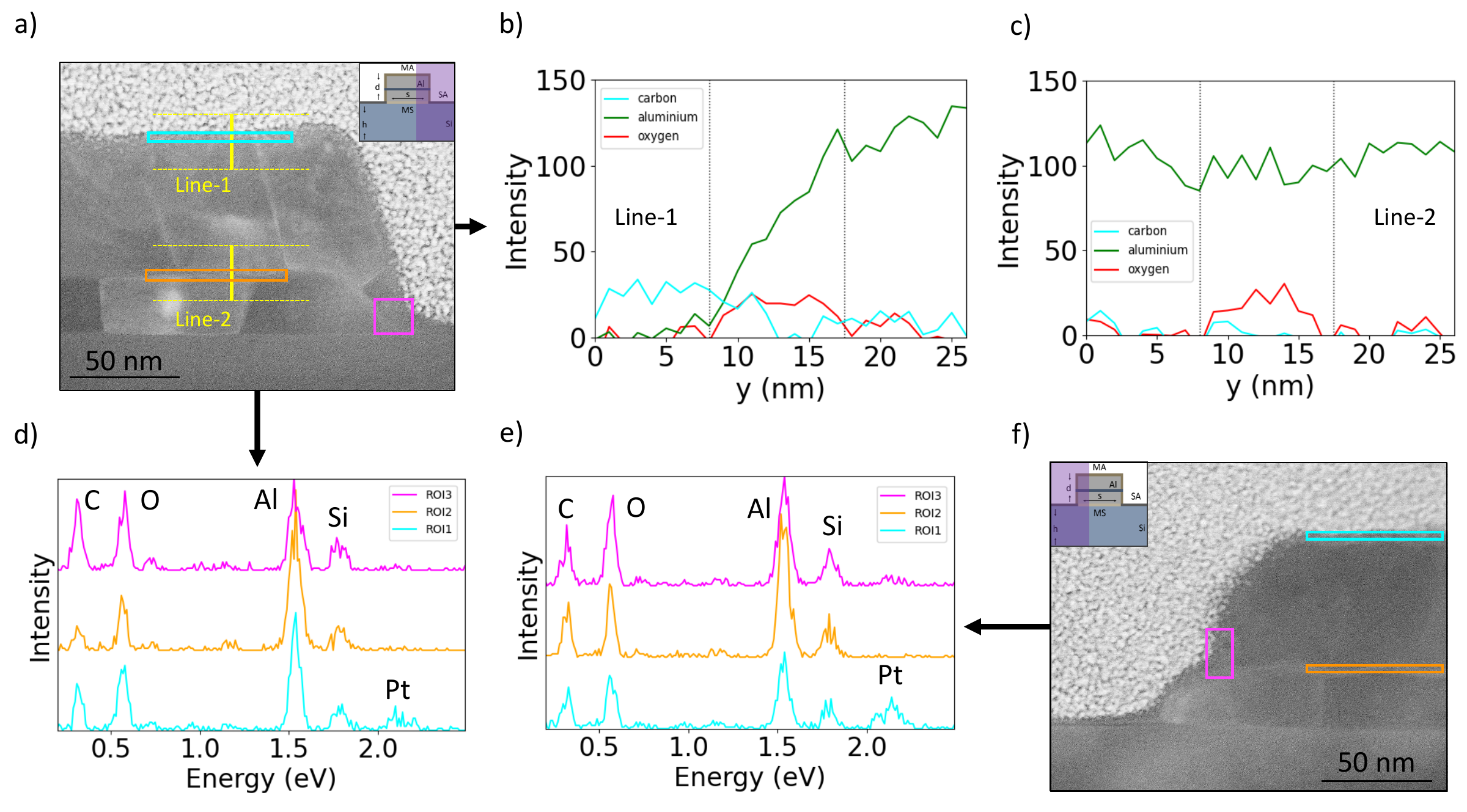}
  \caption{
  Low-Angle Annular Dark Field (LAADF) images in (a) and (f) reveal the grains in the polycrystalline aluminum films due to diffraction contrast. Regions of interest (ROIs) in (d) and (e) are marked: ROI-1 corresponds to the top surface oxide, ROI-2 to the aluminum oxide (AlO$_{\text{x}}$) at the tunnel barrier of the junction, and ROI-3 to the oxide at the side of the junction. Elemental concentrations across these ROIs contain aluminum, oxygen, carbon and platinum. The presence of background silicon is noted throughout the images. Linescans reconstructed from EDX maps in (a) are shown in (b) and (c). (b) Shows elemental profile along Line-1 revealing a surface oxide layer approximately 10~nm thick with increasing aluminum content with depth and carbon pickup at the top of the oxide layer and in the platinum capping layer. (c) Elemental profiles along Line-2 showing a thinner oxide layer about 7~nm thick with lower carbon contamination.}
  \label{FJJ-2}
\end{figure}

The Low-Angle Annular Dark Field (LAADF) images in Fig.~\ref{FJJ-2}(a) and (f) show pronounced diffraction contrast, revealing the grain structure in the polycrystalline aluminum films. Three regions of interest (ROIs) are identified and it is noted that these ROIs include both oxide and some aluminum, making it challenging to isolate the oxide exclusively in these particular maps. Additionally, a background signal from silicon is also present throughout the images. In the right part of the lamella (see Fig.~\ref{FJJ-2}(a) and (d)), we observe that the AlO$_{\text{x}}$ layer within the junction (ROI-2) appears to have the lowest amount of carbon contamination compared to the oxygen-rich layers at the top and side surfaces, suggesting a higher purity in the tunnel barrier. However, in the left section of the lamella  (see Fig.~\ref{FJJ-2}(e) and (f)), we see a similar amount of carbon in all ROIs. Carbon maps show that the platinum coating contains more carbon, but there are patches of carbon in the bulk aluminum area as well. 

The low carbon contamination in the AlO$_{\text{x}}$ tunnel barrier (ROI-2) suggests that the controlled oxidation process effectively produces a high purity insulating layer essential for the junction. However, the pronounced oxygen presence at the MA and SA interfaces indicates that surface oxidation is occurring post-fabrication, likely due to the subsequent fabrication process and exposure to ambient conditions. The carbon levels across different ROIs in the left section of the lamella imply that carbon contamination could be widespread, possibly introduced during sample preparation or from residual processing materials. This pervasive carbon and oxygen at air interfaces could contribute to dielectric loss, building a decoherence channel for the qubit. 

\MA{Both carbon contamination and oxide formation at interfaces potentially originate from several sources, including residual organic films left after lithography and resist stripping, and handling-related contamination introduced during processing and sample preparation \cite{mohseni2024build, zanuz2024mitigating}. These residues contribute to the formation of amorphous dielectric layers, which are known to host TLS defects that can couple to both qubits resonators, acting as parasitic loss channel that absorb and dissipate microwave energy, limiting both qubit coherence times and resonator quality factors \cite{altoe2022localization-27}. Methods to mitigate carbon contamination and TLS loss include optimized resist removal using solvent cleaning combined with in-situ plasma ashing, minimized ambient exposure through vacuum or inert gas transfer between processing steps, and surface passivation immediately after etching to prevent further contamination \cite{alghadeer2022surface-30, alghadeer2024mitigating}. Implementing improved surface cleaning and protective measures during and after the fabrication process could be vital in mitigating these unwanted oxidation and contamination.}

In addition, linescans reconstructed from EDX maps across the surface oxide layers of the Josephson junction area are \MAr{shown in Fig.~\ref{FJJ-2}(b) and (c)}. Line-1 and Line-2 were analyzed to assess the thickness and composition of the oxide layers in different regions. In Line-1, the surface oxide layer is approximately 10~nm thick. The aluminum content appears to increase with depth into the surface; however, here the layers are not completely flat which may cause some smearing of the features. Additionally, there is a noticeable presence of carbon at the very top of the oxide layer within the platinum capping layer, but less within the underlying device. In contrast, Line-2 reveals an oxide layer of about 7~nm thick exhibiting low levels of carbon contamination.

These observations highlight the elemental compositions across different regions of the junction and show the presence of pronounced interface oxidation layers and carbon-based contaminants that may serve as sources of dielectric loss. The thicker oxide layer in Line-1, coupled with the relative carbon presence, suggests that this area may have been more exposed to environmental contaminants or processing residues. The increasing aluminum content with depth indicates a transition from the oxidized surface to the underlying relatively cleaner bulk of metal, but the non-uniformity of the layers necessitates cautious interpretation due to potential smearing effects in the measurement. In contrast, the thinner and cleaner oxide layer observed in Line-2 implies better preservation of the surface with minimal carbon contamination, due to non-exposure to surface treatments and ambient environment. The difference in oxide impurity levels could influence some properties of the junction and contribute to variability in device performance.


\subsection*{Materials Analysis of a Spiral Inductor - Resonator Side}

\begin{figure}[H]
  \centering
   \includegraphics[width=0.9\textwidth]{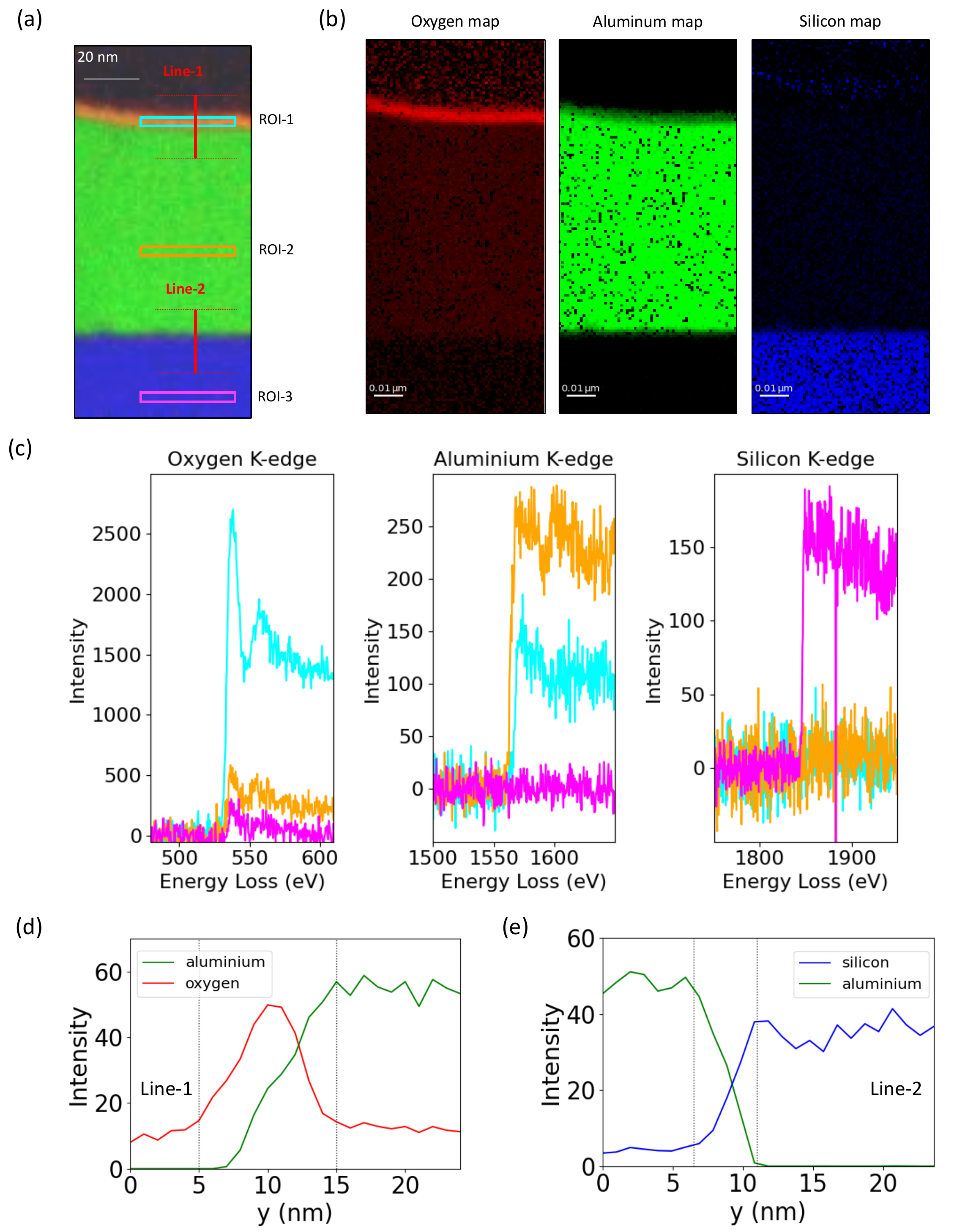}
   \caption{(a) \MA{Scanning transmission electron microscopy (STEM) combined with electron-energy loss spectroscopy (EELS)} maps of the middle section of the lamella extracted from the spiral inductor, with regions of interest across the interface indicated by boxes.\MA{The elemental compositions of oxygen (red), aluminum (green) and silicon (blue) are shown in a combined EELS map}. Individual colored images of (b) oxygen, aluminum, and silicon illustrate high contrast of each elemental distribution. (c) Oxygen K-edge, aluminum K-edge, and silicon K-edge EELS spectra from the three ROIs, showing more oxygen in ROI-1, more aluminum in ROI-2, and more silicon in ROI-3. \MAr{Linescans along different regions in (a) are shown in (d) for Line-1 and (e) Line-2. The elemental profiles along (d) Line-1 reveal a surface oxide layer approximately 10~nm thick, while (e) Line-2 indicates a much cleaner interface with no relative oxygen signal}.}
  \label{FSP-2}
\end{figure}

Next, we discuss materials analysis on the spiral inductor in Fig.~\ref{F1}(b), where the results in Fig.~\ref{FSP-2} presents a detailed analysis of the middle section of the lamella, a key component in the resonator circuit. In Fig.~\ref{FSP-2}(a), the cross-sectional STEM-EELS maps highlight regions of interest (ROIs) across the interface, marked by boxes. The individual images for (b) oxygen, aluminum, and silicon illustrate the distribution of these elements in the examined region. \MAr{Note: the observed silicon signal in the platinum capping layer in (b) is likely an artefact from lamella projection}. The corresponding EELS maps in Figure~\ref{FSP-2}(c) clearly show three distinct regions: AlO$_{x}$ surface (ROI-1), Al film (ROI-2), and Si substrate (ROI-3). In addition, linescans reconstructed from EDX maps along different regions are presented in Figures~\ref{FSP-2}(d) and \ref{FSP-2}(e), corresponding to Line-1, Line-2, and Line-3, respectively. The elemental profiles along Line-1 reveal the presence of a surface oxide layer approximately 10~nm thick, characterized by significant concentrations of aluminum and oxygen. \MAr{In contrast, the profiles along Line-2 show a much cleaner interface with no relative oxygen signal as also seen in Fig.~\ref{FSP-2}(b), suggesting minimal oxidation at this boundary with sharper aluminum and silicon profiles. This observation of a relatively clean MS interface with minimal oxygen content is consistent with the Josephson junction results shown in Fig.~\ref{FJJ-1}(b), where the MS interface was also found to be notably clean across the sample}. These findings highlight the variability in interface quality as the differentiation between oxidized and cleaner regions indicate the importance of precise fabrication techniques to control the oxidation process and maintain the integrity of the interface.

\begin{figure}[H]
  \centering
   \includegraphics[width=0.8\textwidth]{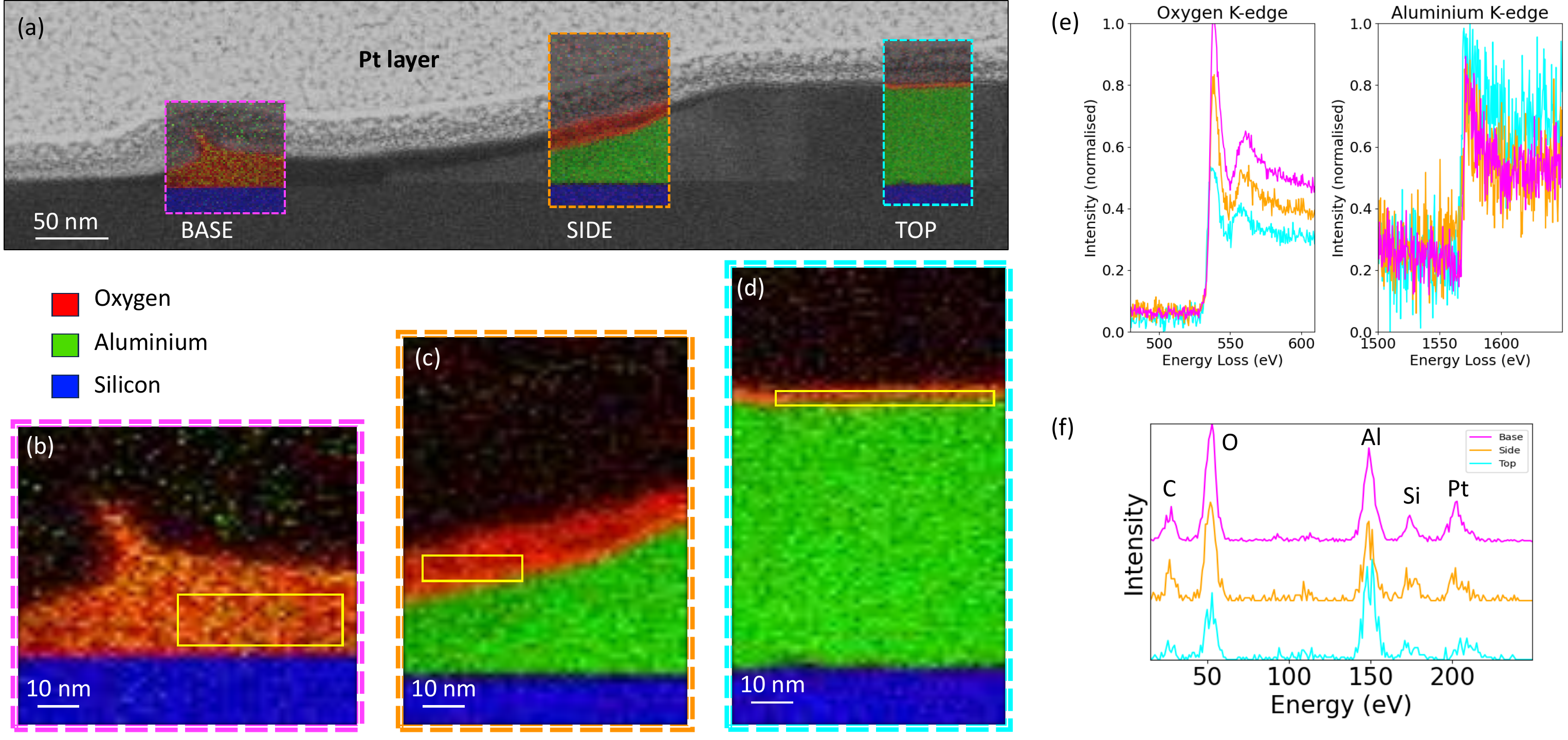}
  \caption{(a) \MA{Cross-sectional scanning transmission electron microscopy (STEM)} image of the left side of the lamella extracted from the resonator side showing locations of \MA{electron-energy loss spectroscopy (EELS)} maps, \MA{highlighting the elemental compositions of oxygen (red), aluminum (green) and silicon (blue)}. EELS maps from (b) the base, (c) the side and (d) the top of the aluminum thin-film, with the region of interest for the AlO\textsubscript{x} spectra indicated as yellow boxes. (e) Oxygen K-edge and aluminum K-edge EELS from the AlO\textsubscript{x} regions. (f) EDX spectra from the same regions of interest.}
  \label{FSP-1}
\end{figure}

Figure~\ref{FSP-1} shows the cross-sectional STEM image of the left side of the cross-section. EELS maps were acquired from three regions of the aluminum thin film: the base, sidewall, and top surface. Analysis of both EELS and EDX data reveals that the oxide layer exhibits increasing oxygen content from the top to the side to the base regions. The aluminum content remains relatively constant throughout these areas. Moreover, EDX results show that the carbon content is higher at the base and sides, suggesting that the thicker oxygen-rich layers observed in these regions may contain a noticeable amounts of organic impurities. In contrast, the oxide layer on the top surface has less carbon, suggesting a cleaner oxide layer.

These observations indicate that the sidewalls and base of the aluminum thin-film are more prone to oxidation and contamination compared to the top surface. The increased oxygen content from top to side to base suggests that these regions accumulate thicker oxide layers. The higher carbon content at the base and sides, as revealed by EDX, implies the presence of organic impurities. This suggests that photolithography and wet-etching steps which selectively target these regions may leave behind contaminants that preferentially adhere to the sidewalls and base. \MAr{While the aluminum intensity appears similar across all three regions (see Fig.~\ref{FSP-1}(a)), the oxygen signal increases progressively from the top to the base. This suggests that the AlO\textsubscript{x} in these areas originates from the same oxidation mechanism during fabrication and ambient exposure, but varies in oxygen level and thickness likely due to spatially non-uniform etching dynamics, particularly at the base}. The cleaner oxide layer on the top surface indicates that this area is less exposed to contaminants or that any impurities are more effectively removed or prevented from adhering during processing. These differences in impurity levels are critical because the thicker, impurity-rich oxide layers at the side and base could introduce additional dielectric loss. Further optimizing the photolithography and etching processes to reduce these imperfections, especially at the sidewalls and base, is essential for enhancing the quality of the processed aluminum thin-film device.

\begin{figure}[H]
  \centering
   \includegraphics[width=0.8\textwidth]{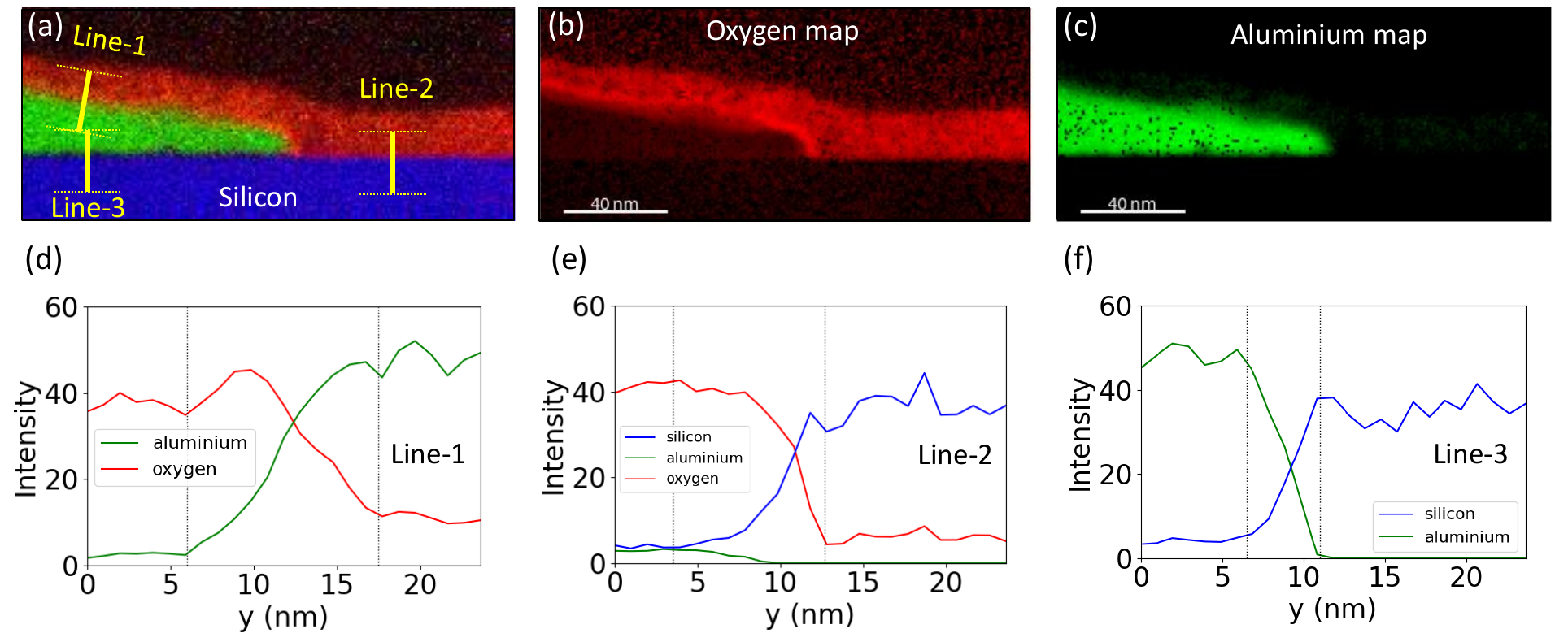}
  \caption{(a) \MA{Cross-sectional scanning transmission electron microscopy (STEM)} images of the right side of the lamella from the resonator side, \MA{highlighting the elemental compositions of oxygen (red), aluminum (green), and silicon (blue)}. (b) Individual colored images of oxygen, and (c) of aluminum, illustrating the presence of aluminum oxide (AlO$_{\text{x}}$). Linescans reconstructed along different regions in (a) are shown in (d) for Line-1, (e) for Line-2, and (f) for Line-3. The elemental profiles along Line-1 show concentrations of aluminum and oxygen, revealing a surface oxide layer and Line-2 show concentrations of silicon and oxygen, while Line-3 indicate a much cleaner interface with no relative oxygen signal.}
  \label{FSP-3}
\end{figure}

Furthermore, Fig.~\ref{FSP-3} provides a detailed compositional analysis of the right side of the cross-section. In Fig.~\ref{FSP-3}(a), the EELS maps highlights the distribution materials composition and the EELS maps in Fig.~\ref{FSP-3}(b) and \ref{FSP-3}(c) display the localized presence of oxygen and aluminum, respectively, confirming the formation of aluminum oxide (AlO$_{\text{x}}$) layers. Linescans reconstructed from EELS maps along three different regions are presented in Fig.~\ref{FSP-3}(d), \ref{FSP-3}(e), and \ref{FSP-3}(f), corresponding to Line-1, Line-2, and Line-3, respectively. The elemental profiles along Line-1 reveal a surface oxide layer, with increasing aluminum content as depth increases, suggesting a gradient in the thickness of the oxide layer from the surface inward. The profiles along Line-2 indicate thicker oxide layer with an increase in silicon content at depth. This may indicate interdiffusion or oxidation extending into both the metal and substrate. In contrast, the profiles along Line-3 show a much cleaner interface, with no relative oxygen signal and sharp transitions in aluminum and silicon content across the interface, suggesting minimal oxidation and a well-defined boundary between the aluminum thin film and the silicon substrate.

\begin{figure}[H]
  \centering
   \includegraphics[width=0.8\textwidth]{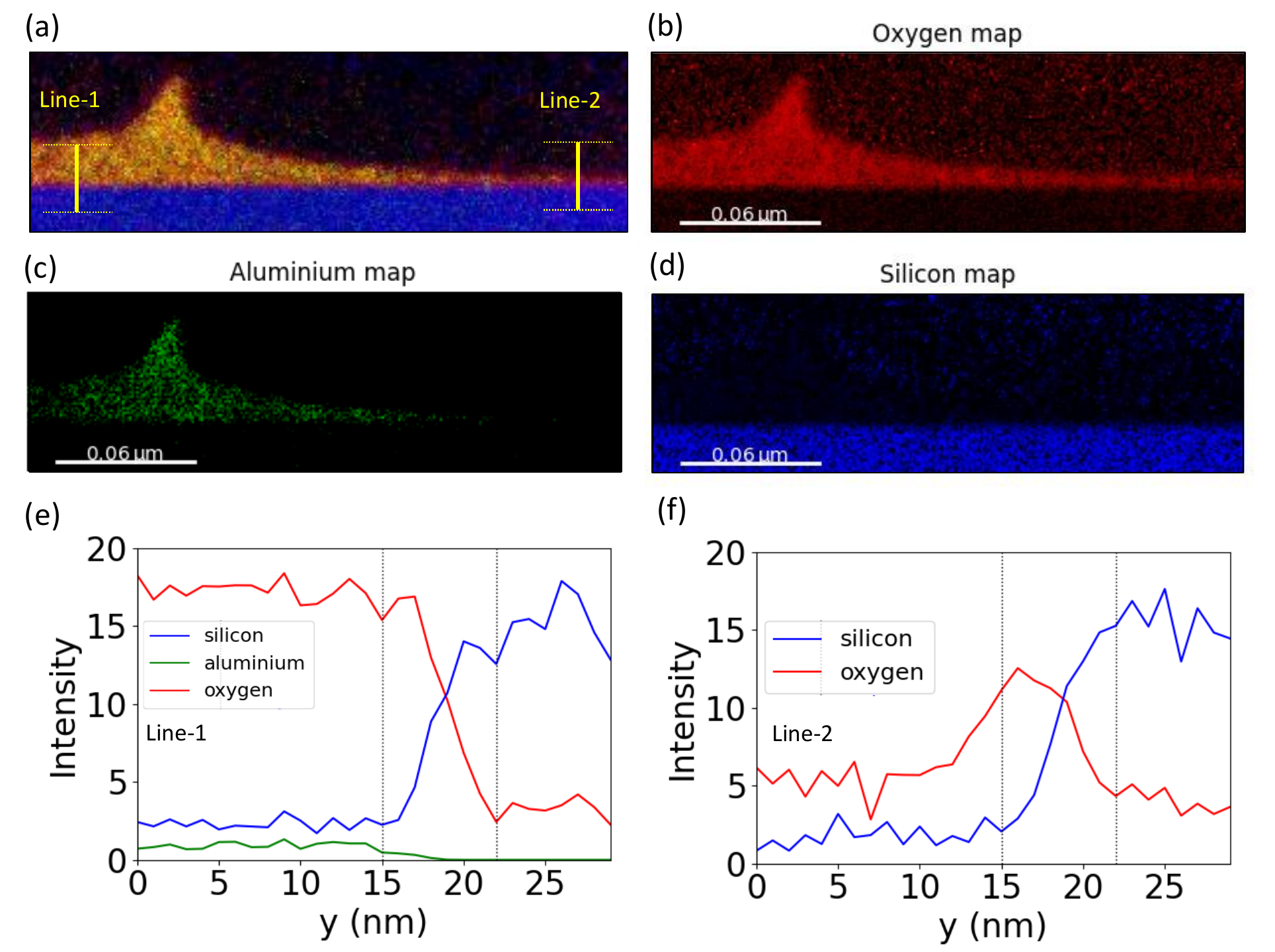}
  \caption{(a) \MA{Cross-sectional scanning transmission electron microscopy (STEM)} colored images at right of the lamella at the location of a structural imperfection from the resonator side, \MA{highlighting the elemental compositions of oxygen (red), aluminum (green), and silicon (blue)}. Individual colored images of the same area for (b) oxygen, (c) aluminum, and (d) silicon, show that this imperfection observed here consists predominantly of aluminum oxide (AlO$_{\text{x}}$). Linescans reconstructed along different regions indicated in (a) are shown in (e) for Line-1 and (f) for Line-2. The elemental profiles along Line-1 display concentrations oxygen, revealing a surface oxide layer approximately 20~nm thick while Line-2, which is further to the right, show non-uniform but thinner oxide layer.}
  \label{FSP-4}
\end{figure}

Additionally, in Fig.~\ref{FSP-4}(a), the cross-sectional STEM images highlight the distribution of materials composition at further right of the lamella. \MAr{The individual elemental maps of the same area confirm that this bulk of the residual layer observed on both sides of the spiral inductor is predominantly composed of oxygen with only a thin presence of aluminum, indicating that it is primarily aluminum oxide (AlO\textsubscript{x}) rather than silicon oxide (SiO\textsubscript{x})}. This indicates that oxidation is occurring mainly within this much thinner aluminum film rather than at the aluminum-silicon interface. From the linescans reconstructed along two different regions in Fig~\ref{FSP-4}(a) we can see the elemental profiles in which Line-1 (see Fig~\ref{FSP-4}(e)) shows concentrations of oxygen, revealing a surface oxide layer approximately 20~nm thick with increasing silicon content as depth increases. This suggests a substantial oxidation layer on the surface in the aluminum thin film. In contrast, the profile along Line-2, which is further to the right, shows non-uniform but thinner oxide layers with increasing silicon content with depth. This may indicate some degree of interfacial oxidation or diffusion of oxygen into the silicon substrate in this region. 

\subsection*{Microwave Characterization}
Basic microwave characterization and device parameters of 16 qubits and 16 resonators from the same wafer are shown in Table. \ref{tab:basic_device_parameters}, using a standard microwave and cryogenic setup described here \cite{rahamim2017double, spring2022high, fasciati2024complementing}. Resonators are designed to have distinct frequencies following a ladder design ranging from \textasciitilde{} 8.6 to 10.0 GHz. Each set of 8 qubits are designed with two-distinct, alternating, frequency pattern in a range between 4.8 and 4.9 GHz, with very low weighted-frequency spreads below 0.5 and 1.5 MHz, receptively, for both targeted values. The average anharmonicity is $\langle \alpha \rangle_{16} = 196.4$ MHz across all qubits with a weighted-frequency spread of 0.3 MHz. The low frequency spreads were achieved without any further post-fabrication process on the junctions, such as junctions annealing \cite{hertzberg2021laser}, but only by fine tuning the junctions fabrication parameters.

Qubits relaxation and coherence times were measured repeatedly over 12 hours, resulting in a total of 400 measurements for each $T_{1}$, $T_{2R}$ and $T_{2E}$ that are then averaged for each qubit. The average qubits relaxation times $T_1$, and dephasing times $T_2R$ and $T_{2E}$ were measured using Ramsey and Hahn echo sequences, respectively. Relaxation and coherence times averaged across the device are $\langle T_{1} \rangle_{16} = 71 ~\pm ~5 ~\mu s$, $\langle T_{2R} \rangle_{16} = 51 ~\pm ~4 ~\mu s$ and $\langle T_{2E} \rangle_{16} = 78 ~\pm ~5 ~\mu s$, with weighted-standard deviations.

\MAr{While a direct quantitative correlation between atomic-scale material imperfections and individual qubit performance is not established here, the inclusion of microwave characterization provides a meaningful benchmark for circuits fabricated using the same process flow analyzed in this study. The coherence times and quality factors reported across the 16 qubits and resonators offer insight into the level of performance achievable in the presence of the types of structural imperfections observed using the presented materials characterization techniques. We emphasize that although localized defects (such as interface oxidation or carbon contamination) may vary in position and extent, the fabrication process is consistent across the device, allowing us to infer a representative range of performance outcomes resulting from these imperfections.}

\begin{table}[htbp]
    \centering
    \caption{\MA{Basic Device Parameters and microwave characterization. \MAr{\( w_{r}/2\pi \) and \( w_{q}/2\pi \) are frequencies of the readout resonator and qubit, respectively. \( Q_{i} \) is the internal quality factor of the resonator, and \( \kappa_{\text{ext}} \) is the external coupling rate. \( \chi \) is the qubit-resonator dispersive shift, and \( \alpha \) is the qubit anharmonicity}. Qubits relaxation and coherence times $T_{1}$, $T_{2R}$ and $T_{2E}$ are averaged over 400 repeated measurements.}}
    \resizebox{1.0\textwidth}{!}{ 
        \begin{tabular}{|c|c|c|c|c|c|c|c|c|c|}
            \hline
            Parameters & \( w_{r}/2\pi \) & \( w_{q}/2\pi \) & \( Q_{i} \) & \( k_{ext} \) & \( \chi \) & \( \alpha \)  & \( \langle T_{1} \rangle \) & \( \langle T_{2R} \rangle \) & \( \langle T_{2E} \rangle \) \\
            \hline
            Qubits & MHz & MHz & \( 10^{4} \) & MHz & KHz  & MHz & \( \mu s \) & \( \mu s \) & \( \mu s \) \\
            \hline
            \( Q_{1} \)  & 9997.4 & 4888.2 & 11.7 & 2.6 & -200.0 & -196.6 & 126 ± 18 & 107 ± 12 & 124 ± 23 \\
            \( Q_{2} \)  & 9386.0 & 4795.6 & 11.8 & 1.3 & -225.0 & -197.2 & 89 ± 13  & 56 ± 15  & 86 ± 12 \\
            \( Q_{3} \)  & 9299.2 & 4807.5 & 6.5  & 1.8 & -200.0 & -196.2 & 61 ± 6   & 44 ± 5   & 102 ± 18 \\
            \( Q_{4} \)  & 8649.5 & 4809.8 & 5.3  & 2.8 & -200.0 & -198.6 & 54 ± 9   & 39 ± 14  & 97 ± 20 \\
            \( Q_{5} \)  & 8755.6 & 4855.3 & 3.1  & 0.8 & -225.0 & -196.4 & 68 ± 10  & 38 ± 4   & 45 ± 10 \\
            \( Q_{6} \)  & 9220.6 & 4824.8 & 6.4  & 0.5 & -225.0 & -194.0 & 63 ± 7   & 49 ± 4   & 68 ± 10 \\
            \( Q_{7} \)  & 9474.2 & 4928.5 & 10.8 & 1.7 & -175.0 & -195.6 & 77 ± 12  & 45 ± 12  & 82 ± 15 \\
            \( Q_{8} \)  & 9908.6 & 4829.5 & 5.8  & 2.0 & -175.0 & -197.2 & 63 ± 8   & 32 ± 7   & 71 ± 8 \\
            \( Q_{9} \)  & 9802.3 & 4963.4 & 7.8  & 4.4 & -250.0 & -195.0 & 24 ± 7   & 24 ± 5   & 33 ± 9 \\
            \( Q_{10} \) & 9535.4 & 4817.2 & 7.3  & 3.2 & -200.0 & -196.9 & 51 ± 10  & 35 ± 9   & 63 ± 17 \\
            \( Q_{11} \) & 9112.8 & 4835.4 & 16.5 & 1.4 & -175.0 & -196.1 & 74 ± 7   & 49 ± 3   & 78 ± 13 \\
            \( Q_{12} \) & 8851.8 & 4777.3 & 11.4 & 0.9 & -250.0 & -196.4 & 92 ± 24  & 57 ± 11  & 76 ± 15 \\
            \( Q_{13} \) & 8943.2 & 4884.0 & 16.7 & 1.6 & -175.0 & -195.3 & 102 ± 13 & 63 ± 7   & 107 ± 22 \\
            \( Q_{14} \) & 9025.3 & 4855.2 & 9.3  & 1.4 & -175.0 & -197.0 & 56 ± 12  & 56 ± 10  & 60 ± 15 \\
            \( Q_{15} \) & 9645.5 & 5040.2 & 7.3  & 1.7 & -225.0 & -196.1 & 55 ± 7   & 58 ± 9   & 65 ± 9 \\
            \( Q_{16} \) & 9728.7 & 4792.8 & 24.6 & 2.7 & -175.0 & -197.5 & 60 ± 6   & 46 ± 4   & 65 ± 9 \\
            \hline
            Statistics  \\
            \hline
            \( \text{Max} \)                   & 9997.4 & 5040.2 & 24.6 & 4.4  & -175.0 & -194.0 & 126 & 107 & 124 \\
            \( \text{Min} \)                   & 8649.5 & 4777.3 & 3.1  & 0.5  & -250.0 & -198.6 & 41  & 32  & 45 \\
            \( \mu \) (Mean)                   & 9333.5 & 4856.5 & 10.1 & 1.9  & -203.1 & -196.4 & 71  & 51  & 78 \\
            \( \sigma \) (Std. Dev)            & 407.12 &   -    & 5.3  & 0.96 & 26.3   & 1.1    & 21  & 17  & 20 \\
            \( \sigma/\sqrt{N} \) (\( N=16 \)) & 101.8  &   -    & 1.3  & 0.2  & 6.6    & 0.3    & 5   & 4   & 5 \\
            \hline
        \end{tabular}
    }
    \label{tab:basic_device_parameters}
\end{table}

\section*{Conclusion}
In this work, we present atomic-level characterization of materials imperfections on a superconducting quantum device. Our findings highlight the presence of pronounced surface and interface oxidation layers and carbon-based contaminants that may serve as sources of dielectric loss, contributing to decoherence in superconducting qubits and resonators. Notably, we observed significant non-uniformity in oxide layers thickness and impurity levels across different interfaces. High-resolution STEM, EDS, and EELS analyses demonstrated pronounced oxidation and carbon contamination at the metal-air (MA) and substrate-air (SA) interfaces, while the metal-substrate (MS) interface remained relatively clean on both sides of the device. These findings emphasize on the fact that surface oxidation and organic residues - likely introduced during fabrication and environmental exposure - are critical sources of materials loss. 

\section*{Acknowledgment}
This work has received funding from the United Kingdom Engineering and Physical Sciences Research Council (EPSRC) under Grants No. EP/N015118/1 and No. EP/T001062/1. M.B. acknowledges support from EPSRC QT Fellowship under Grant No. EP/W027992/1. S.C. acknowledges support from Schmidt Science. 

We thank Diamond Light Source for access and support in use of the electron Physical Science Imaging Centre (Instrument E01 and E02 and proposal number MG37092) that contributed to the results presented here. We would like to acknowledge the Superfab Nanofabrication facility at Royal Holloway, University of London, and Optoelectronics Research Centre at University of Southampton where part of device fabrication was performed.




\bibliography{Biography}

\end{document}